\documentclass[onecolumn]{emulateapj}

\slugcomment{}

\shorttitle{Giant Broad Line Regions}
\shortauthors{Devereux et al.}

\begin{document}
\columnsep=8.5mm
\textwidth=18.55cm
\title{Giant Broad Line Regions in Dwarf Seyferts}

\author{Nick Devereux}
\affil{Department of Physics \& Astronomy, Embry-Riddle Aeronautical University,
         Prescott, AZ 86301: devereux@erau.edu}

\begin{abstract}

High angular resolution spectroscopy obtained with the Hubble Space Telescope ({\it HST}) has revealed a remarkable population of galaxies hosting dwarf Seyfert nuclei with an unusually large broad-line region (BLR). These objects are remarkable for two reasons. Firstly, the size of the BLR can, in some cases, rival those seen in the most luminous quasars. Secondly, the size of the BLR is not correlated with the central continuum luminosity, an observation that distinguishes them from their reverberating counterparts. Collectively, these early results suggest that non-reverberating dwarf Seyferts are a heterogeneous group and not simply scaled versions of each other. Careful inspection reveals broad H Balmer emission lines with single peaks, double peaks, and a combination of the two, suggesting that the broad emission lines are produced in kinematically distinct regions centered on the black hole (BH). Because the gravitational field strength is already known for these objects, by virtue of knowing their BH mass, the relationship between velocity and radius may be established, given a kinematic model for the BLR gas. In this way, one can determine the inner and outer radii of the BLRs by modeling the shape of their broad emission line profiles. In the present contribution, high quality spectra obtained with the Space Telescope Imaging Spectrograph (STIS) are used to constrain the size of the BLR in the dwarf Seyfert nuclei of M81, NGC 3998, NGC 4203, NGC 3227, NGC 4051, and NGC 3516.

 \end{abstract}


\section{Introduction}

The physics behind the production of broad emission lines in active galactic nuclei (AGN) has challenged generations of astronomers since the phenomenon was first reported in a seminal paper by \cite{Sey43}. Much has been learned over the intervening years, primarily as a result of reverberation mapping \citep{Pet93}, which has led to an estimate for the size of the BLR producing the broad Balmer emission lines. For reverberating AGN, the broad Hβ emission line flux varies typically by ${\sim}$15\% in a correlated response to larger ${\sim}$30\% changes in the level of the adjacent continuum. The light-travel-time delay between the brightening of the broad Hβ emission line, following a brightening of the continuum, leads to an estimate for the “size” of the BLR. There is some difference of opinion over what size is actually being measured. One possibility is that the size is a luminosity-weighted radius as proposed by \cite{Kor04}. Another possibility is that it is the inner radius of a much larger region of ionized gas, as suggested most recently by \cite{DH13}.  

Regardless of the definition of the word``size", there is no question that a superb correlation exists between the size and the continuum luminosity for reverberating AGN, a correlation characterized in detail by \cite{Kas05}. That correlation shows that the smallest BLR occurs in the lowest luminosity reverberating AGN. It thus came as a complete surprise when using a different technique, line profile fitting, an enormous size was inferred for the BLR of the low luminosity active galactic nuclei (LLAGN) M81 \citep{Dev07}, NGC 3998 \citep{Dev11a}, and NGC 4203 \citep{Dev11b}. Similarly large BLR sizes have been inferred for other LLAGN, using different analysis techniques, by \cite{Wan03}, \cite{Zha07}, and \cite{Bal14}. Collectively, these studies lead inevitably to the strange and unexpected conclusion that many LLAGN, the so-called dwarf Seyferts \citep{Ho97a,Ho97b}, possess giant BLRs. 

\section{The Line Profile Fitting Method for Determining BLR Size}

The line profile fitting method for estimating the size of the region poducing broad Balmer line emission
has been described previously \citep{Dev11a}. Briefly, the method employs a Monte Carlo simulation of 
${\sim}$ 10$^4$ particles of light moving under the influence of gravity in free--fall according to the familiar equation v(r) = $\rm {\sqrt{ 2 G M_{\bullet}/r}}$, where v is velocity, G is the gravitational constant, M$_{\bullet}$ is the BH mass, and r is the distance of each point from the central supermassive BH. Discrete particle models have the advantage that they reproduce the small-scale structure seen in broad emission line profiles, which is caused by random clumping in radial velocity space, as noted previously by \cite{Cap81}.

The model produces a single peak broad emission line profile for a spherically symmetric distribution of points organized according to a power law, $\rho(r) \propto r{^{-n}}$. The index value {\it n} = 1.5 
is special as it produces the line profile shape expected for a steady-state inflow\footnotemark \footnotetext[1]{Or outflow, the only difference being a minus sign representing the direction of the velocity vector.} of points.
In this case, there are only two free parameters; the inner and outer radii of the volume containing the model 
points, which can be varied to produce a wide variety of model broad emission line profile shapes. The inner radius of the volume defines the full velocity width at zero intensity of the model emission line, and the outer radius of the volume defines the maximum intensity of the model emission line at zero velocity. Thus, 
comparing a normalized version of the observed broad emission line with the model one effectively constrains the inner and outer radii of the emitting volume with high precision using chi-squared minimization. A merit of applying the line profile fitting method to LLAGNs is that they radiate well below the Eddington limit and are therefore unable to sustain a radiatively driven outflow. Consequently, gravity dominates the kinematics of the BLR gas, and the three dimensional gravitational field strength is known by virtue of knowing the central mass. 

\begin{deluxetable*}{lccclccc}
\tabletypesize{\scriptsize}
\tablecolumns{8} 
\tablecaption{Non-Reverberating \& Reverberating Dwarf Seyferts\label{tbl-2}}
\tablewidth{0pt}
\tablehead{
\colhead{Object} & \colhead{Morphology} & \colhead{Distance} & \colhead{Spectral\tablenotemark{a}} &
\colhead{Object} & \colhead{Morphology} & \colhead{Distance} & \colhead{Spectral\tablenotemark{a}} \\
& & \colhead{(Mpc)} & \colhead{Classification}  & & & \colhead{(Mpc)} & \colhead{Classification}  \\
\colhead{(1)} & \colhead{(2)} &  \colhead{(3)} & \colhead{(4)}  & \colhead{(5)} & \colhead{(6)} &  \colhead{(7)} & \colhead{(8)}  \\
}
\startdata
NGC 3031 (M81) & SA(s)ab & 3.6  & S1.5 & NGC 3227  & SAB(s)a & 20.8  & S1.5 \\
NGC 3998 & SA0 & 14.1 & L1.9 & NGC 3516 & SB0 & 38.0 & S1.2 \\ 
NGC 4203 & SAB0 & 15.1 &  L1.9 & NGC 4051 & SBbc & 17.9 & S1.2 \\

\enddata

\tablenotetext{a}{\cite{Ho97a}}

\end{deluxetable*}

\section{Results}

There are presently six LLAGNs whose BLR size has been deduced by modeling the shape of the H Balmer lines  including three non-reverberating dwarf Seyferts (M81, NGC 3998 and NGC 4203), and three reverberating ones (NGC 3227, NGC 3516 and NGC 4051).  Some basic observables are listed
in Table 1, and their BLR size-luminosity relationship is illustrated in Figure 1. 
The three
non-reverberating dwarf Seyferts lie far from the correlation defined by reverberating AGNs. The other three LLAGN are reverberating, and their inner radii consistently lie on the correlation defined by other reverberating AGN. This latter result is important as it shows that the line profile fitting method does yield a size that agrees with the reverberation size. In particular, the inner radius determined from line profile fitting coincides almost identically with the reverberation radius as demonstrated for NGC 3227 \citep{Dev13}, NGC 4051 \citep{DH13} and NGC 3516 (Devereux 2015, in prep). This coincidence between the reverberation radius and the inner radius determined from line profile fitting can be understood if both are measuring the gas closest to the BH. With hindsight, this coincidence is perhaps not too surprising as both reverberation mapping and line profile fitting employ the familiar velocity equation v(r) = $\rm {\sqrt{\gamma G M_{\bullet}/r}}$, where ${\gamma}$ is a constant describing the gas kinematics. However, by virtue of utilizing the entire broad emission line flux, not just the ${\sim}$ 15\% or so that is reverberating, line profile fitting yields the full extent of the BLR with greater efficacy than reverberation mapping. Consequently, one discovers that the outer radius of the BLR is quite large even for the reverberating dwarf Seyferts but not as large as for the non-reverberating ones, which, by comparison, have a giant broad line region.

\begin{figure}
\epsscale{1.0}
\centering
\includegraphics[width=84mm,clip]{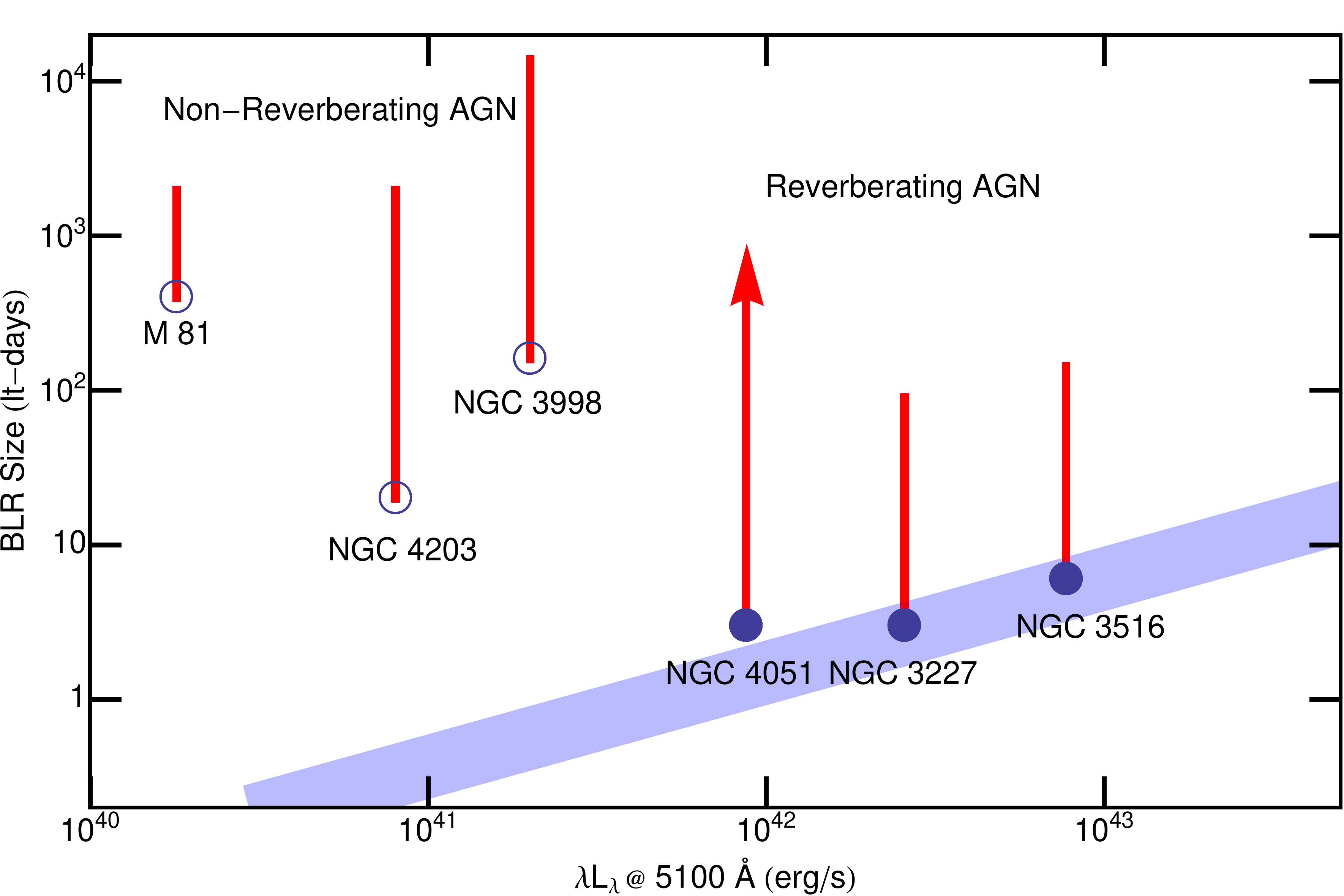}
\caption{Broad line region size in light-days versus the continuum luminosity at 5100 {\AA} in erg/s.  Open circles identify the non-reverberating dwarf Seyferts with a giant broad line region (M81, NGC 3998 and NGC 4203). Filled circles identify the reverberating dwarf Seyferts (NGC 4051, NGC 3227 and NGC 3516). Vertical red lines connect the inner and outer radii of the BLR determined using line profile fitting. The arrow indicates that only a lower limit could be determined for the outer radius of the BLR in NGC 4051. For the reverberating dwarf Seyferts the inner radius coincides with the reverberation radius. The correlation between reverberation radius and luminosity established for reverberating AGN is identified by the blue-shaded band. Even the reverberating dwarf Seyferts have large outer radii, but much smaller than the non-reverberating dwarf Seyferts which have giant broad line regions. }
\end{figure}

An illustration of the relative sizes of the BLR, on a linear scale, is provided in Figure 2. 
The angular size of the BLR in 
NGC 3998 and M81 is so large that it is partially obscured by the 0.1${\arcsec}$ slit employed for the STIS observations. NGC 4051 is a narrow line Seyfert 1 for which the line profile fitting method yields only a lower limit to the outer radius as discussed previously \citep{DH13}.

\section{Discussion}

The increased scatter in the size-luminosity relationship for LLAGNs illustrated in Figure 1 is not to be confused with the contraindication of low scatter in the size-luminosity correlation discussed recently by \cite{Kil15}. Evidently, there are two AGN populations that have very different size-luminosity relations. Firstly, there are the well known reverberating AGN which define the size-luminosity correlation. Then there are the other, lesser known but more populous, LLAGN that are not reverberating, and for which the BLR size is evidently not correlated with AGN continuum luminosity. It is this latter, rather neglected, but larger segment of the LLAGN population, that is the main focus of this paper. Collectively, LLAGN exhibit no correlation between size and luminosity as illustrated in Figure 1. 
Consequently, it is inappropriate to estimate BH masses for non-reverberating AGN using the size-luminosity relationship established for reverberating AGN \cite[e.g.,][]{Gre05,Gre07}. 

Non-reverberating LLAGN are distinguished from their reverberating counterparts in four ways. First, the size of the BLR is much larger. Second, the size of the BLR is not correlated with the AGN continuum luminosity indicating that they are not simply scaled versions of each other.  Third, even though in some cases the UV continuum
may be time variable \cite{Mao05}, there is as yet no correlated response measured for the Balmer emission lines. Fourth, at least two objects M81 and NGC 4203 exhibit time variable, double-peak broad Balmer emission lines \citep{Dev03,Dev11b}. For the latter object, the Balmer line fluxes increased by a factor of two over a time interval of a decade with no perceptible
change in the adjacent continuum. Such emission line variability is an order of magnitude larger than seen among the reverberating population. 

Figure 2 graphically illustrates that the BLR in the non-reverberating dwarf Seyfert NGC 3998 is by far the largest of the six objects included in this study. It is also the first AGN to be identified with broad forbidden lines including [O\,{\sc iii}]${\lambda\lambda}$4959,5007 and [S\,{\sc ii}]${\lambda\lambda}$6718,6732. That the forbidden [S\,{\sc ii}] lines have the same width as the permitted H Balmer lines allows a useful limit to be placed on the average electron density of the BLR gas corresponding to ${\sim}$ 7 ${\times}$ 10${^3}$ cm$^{-3}$. As explained in more detail by \cite{Dev11a}, the BLR in NGC 3998 is most easily explained in terms of an ionization bounded, low density inflow of H gas that is photoionized by the central UV-X--ray source.

One property common to both the reverberating and non-reverberating dwarf Seyferts is the existence of a 
central hole or cavity in the photoionized H gas, evidenced by the finite width at zero intensity of the H
Balmer lines. The apparent absence of Balmer line emission inside the inner radius, synonymous now with the reverberation radius, is most easily explained if the electron temperature exceeds ${\sim}$ 10${^6}$ K, in which case the primary source of opacity is electron scattering.

\begin{figure}
\epsscale{1.0}
\centering
\includegraphics[width=180mm,clip]{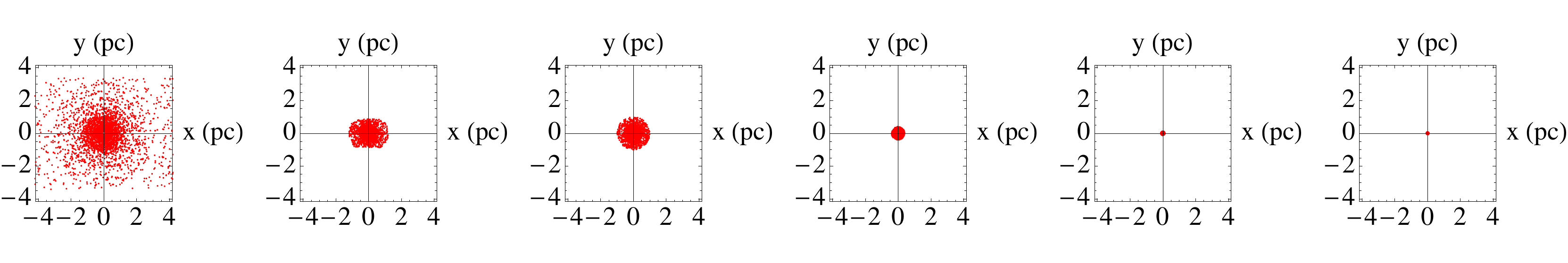}
\caption{Comparison of BLR physical size in pc.  From left to right: NGC 3998, M81, NGC 4203, NGC 4051, NGC 3516 and NGC 3227. The large angular size of the BLR in NGC 3998 and M81 caused the STIS spectrometer slit to occult some of the light. Thus, the shape of the BLR appears to be flattened in these two dwarf Seyferts.  The outer radius of the BLR in NGC 4051 is a lower limit (see text for details).}
\end{figure}

\section{Conclusions}

High angular resolution spectroscopic observations obtained with {\it HST} have revealed a previously unrecognized population of
non-reverberating dwarf Seyferts with a giant broad-line region. 
The existence of these objects significantly diminishes the correlation between the central continuum luminosity and BLR size among LLAGN. With a diameter of ${\sim}$ 14 pc, the BLR in NGC 3998 is the largest recorded to date. However, the size of the BLR remains to be established for approximately 20 additional non-reverberating dwarf Seyferts residing in the local universe.


\begin{thebibliography}{99}

\bibitem[Balmaverde \& Capetti (2014)]{Bal14} Balmaverde, B., \& Capetti, A., 2014, \aap, 563, 119

\bibitem[Capriotti, Foltz \& Byard (1981)]{Cap81} Capriotti, E., Foltz, C., \& Byard, P., 1980, \apj, 245, 396

\bibitem[Devereux (2013)]{Dev13} Devereux, N., 2013, \apj, 764, 79

\bibitem[Devereux (2011a)] {Dev11a} Devereux, N., 2011, \apj, 727, 93

\bibitem[Devereux (2011b)] {Dev11b} Devereux, N., 2011, \apj, 743, 83

\bibitem[Devereux et al.(2003)]{Dev03} Devereux, N., Ford, H., Tsvetanov, Z., \&   Jacoby, G., 2003, \apj, 125, 1226  

\bibitem[Devereux \& Heaton (2013)]{DH13} Devereux, N., \& Heaton, E., 2013, \apj, 773, 97

\bibitem[Devereux \& Shearer (2007)]{Dev07} Devereux, N.,  \&   Shearer, A., 2007, \apj, 671, 118

\bibitem[Greene \& Ho (2005)]{Gre05} Greene, J.E., \& Ho, L.C., 2005, \apj, 630, 122

\bibitem[Greene \& Ho (2007)]{Gre07} Greene, J.E., \& Ho, L.C., 2007, \apj, 667, 131

\bibitem[Ho, Filippenko \& Sargent (1997)]{Ho97a} Ho, L. C., Filippenko, A. V.,  \& Sargent, W. L. W.,  1997, \apjs, 112, 315

\bibitem[Ho et al. (1997)]{Ho97b} Ho, L. C., Filippenko, A. V., Sargent, W.L.W., \& Peng, C. Y.,  1997, \apjs, 112, 391

\bibitem[Korista \& Goad (2004)]{Kor04} Korista, K.T. \& Goad, M.R., 2004, \apj, 606, 749

\bibitem[Kaspi et al. (2005)]{Kas05} Kaspi, S., Maoz, D., Netzer, H., et al. 2005, \apj, 629, 61

\bibitem[Kilerci Eser et al. (2015)]{Kil15} Kilerci Eser, E., Vestergaard, M., Peterson, B. M., et al. 2015, \apj, 801, 8

\bibitem[Maoz et al., (2005)]{Mao05} Maoz, D., Nagar, N.M., Falcke, H., \& Wilson, A., 2004, \apj, 625, 699

\bibitem[Peterson (1993)]{Pet93} Peterson, B., 1993, \pasp, 105, 247

\bibitem[Seyfert (1943)]{Sey43} Seyfert. C.K., 1943, \apj, 97, 28 

\bibitem[Wang \& Zhang (2003)]{Wan03}    Wang, T-G \& Zhang, X-G, 2003, \mnras, 340, 793	

\bibitem[Zhang, Dultzin-Hacyan \& Wang (2007)]{Zha07} Zhang, X-G; Dultzin-Hacyan, D., \& Wang, T-G., 2007, \mnras, 374, 691



\end{thebibliography}
\end{document}